\documentclass[floatfix,twocolumn,showpacs,preprintnumbers,amsmath,amssymb,pra,superscriptaddress,longbibliography]{revtex4-1}
\usepackage{color}
\usepackage[usenames,dvipsnames,svgnames,table]{xcolor}
\usepackage[colorlinks=true,linkcolor=blue,urlcolor=blue,citecolor=blue]{hyperref}
\usepackage{mathtools}
\usepackage{graphicx}
\usepackage{dcolumn}
\usepackage{array}
\usepackage{lipsum}
\usepackage{bm}
\usepackage{subfigure}
\usepackage{amssymb}
\usepackage{multirow}
\usepackage{tabularx}
\usepackage{amsmath}
\usepackage{braket}
\usepackage{csquotes}
\graphicspath{{plots/}}
 \usepackage{lipsum}
\usepackage{mathrsfs}

\newcommand{\beq}{\begin{equation}}
\newcommand{\eeq}{\end{equation}}
\newcommand{\bea}{\begin{eqnarray}}
\newcommand{\eea}{\end{eqnarray}}

\begin{document}
\title{Model-free Rayleigh weight from x-ray Thomson scattering measurements}

\author{T.~Dornheim}
\email{t.dornheim@hzdr.de}

\affiliation{Center for Advanced Systems Understanding (CASUS), 02826 G\"orlitz, Germany}
\affiliation{Helmholtz-Zentrum Dresden-Rossendorf (HZDR), 01328 Dresden, Germany}

\author{H.~M.~Bellenbaum}

\affiliation{Center for Advanced Systems Understanding (CASUS), 02826 G\"orlitz, Germany}
\affiliation{Helmholtz-Zentrum Dresden-Rossendorf (HZDR), 01328 Dresden, Germany}
\affiliation{Institut f\"ur Physik, Universit\"at Rostock, 18051 Rostock, Germany}
\affiliation{Lawrence Livermore National Laboratory (LLNL), California 94550 Livermore, USA}

\author{M.~Bethkenhagen}
\affiliation{LULI, CNRS, CEA, Sorbonne Universit\'{e}, \'{E}cole Polytechnique – Institut Polytechnique de Paris, 91128 Palaiseau, France}


\author{S.~B.~Hansen}
\affiliation{Sandia National Laboratories, Albuquerque, NM 87185, USA}

\author{M.~P.~B\"ohme}
\affiliation{Lawrence Livermore National Laboratory (LLNL), California 94550 Livermore, USA}
\affiliation{Center for Advanced Systems Understanding (CASUS), 02826 G\"orlitz, Germany}
\affiliation{Helmholtz-Zentrum Dresden-Rossendorf (HZDR), 01328 Dresden, Germany}

\author{T.~D\"oppner}
\affiliation{Lawrence Livermore National Laboratory (LLNL), California 94550 Livermore, USA}

\author{L.~B.~Fletcher}
\affiliation{SLAC National Accelerator Laboratory, Menlo Park, California 94309, USA}

\author{Th.~Gawne}
\affiliation{Center for Advanced Systems Understanding (CASUS), 02826 G\"orlitz, Germany}
\affiliation{Helmholtz-Zentrum Dresden-Rossendorf (HZDR), 01328 Dresden, Germany}

\author{D.~O.~Gericke}
\affiliation{Centre for Fusion, Space and Astrophysics, University of Warwick, Coventry CV4 7AL, UK}

\author{S.~Hamel}
\affiliation{Lawrence Livermore National Laboratory (LLNL), California 94550 Livermore, USA}

\author{D.~Kraus}
\affiliation{Institut f\"ur Physik, Universit\"at Rostock, 18051 Rostock, Germany}
\affiliation{Helmholtz-Zentrum Dresden-Rossendorf (HZDR), 01328 Dresden, Germany}

\author{M.~J.~MacDonald}
\affiliation{Lawrence Livermore National Laboratory (LLNL), California 94550 Livermore, USA}

\author{Zh.~A.~Moldabekov}
\affiliation{Center for Advanced Systems Understanding (CASUS), 02826 G\"orlitz, Germany}
\affiliation{Helmholtz-Zentrum Dresden-Rossendorf (HZDR), 01328 Dresden, Germany}

\author{Th.~R.~Preston}
\affiliation{European XFEL, 22869 Schenefeld, Germany}

\author{R.~Redmer}
\affiliation{Institut f\"ur Physik, Universit\"at Rostock, 18051 Rostock, Germany}

\author{M.~Sch\"orner}
\affiliation{Institut f\"ur Physik, Universit\"at Rostock, 18051 Rostock, Germany}

\author{S.~Schwalbe}
\affiliation{Center for Advanced Systems Understanding (CASUS), 02826 G\"orlitz, Germany}
\affiliation{Helmholtz-Zentrum Dresden-Rossendorf (HZDR), 01328 Dresden, Germany}

\author{P.~Tolias}
\affiliation{Space and Plasma Physics, Royal Institute of Technology (KTH), Stockholm, SE-100 44, Sweden}

\author{J.~Vorberger}
\affiliation{Helmholtz-Zentrum Dresden-Rossendorf (HZDR), 01328 Dresden, Germany}

\begin{abstract}
X-ray Thomson scattering (XRTS) has emerged as a powerful tool for the diagnostics of matter under extreme conditions. In principle, it gives one access to important system parameters such as the temperature, density, and ionization state, but the interpretation of the measured XRTS intensity usually relies on theoretical models and approximations. 
In this context, a key property is given by the Rayleigh weight that describes the electronic localization around the ions.
Here, we show that it is possible to extract the Rayleigh weight directly from the experimental data without the need for any model calculations or simulations. As a practical application, we consider an experimental measurement of strongly compressed Be at the National Ignition Facility (NIF) [D\"oppner \emph{et al.}, \textit{Nature} \textbf{618}, 270-275 (2023)]. 
We demonstrate that experimental results for the Rayleigh weight open up new avenues for the interpretation of XRTS experiments by matching the measurement with \emph{ab initio} simulations such as density functional theory or path integral Monte Carlo. Interestingly, this new procedure leads to significantly lower density compared to previously used chemical models.
\end{abstract}

\maketitle

The x-ray Thomson scattering (XRTS) method~\cite{siegfried_review,sheffield2010plasma} constitutes a powerful experimental technique, which is capable of giving microscopic insights into a probed sample.
A particularly important use case for XRTS is the diagnostics of experiments with matter under extreme densities, temperatures and pressures~\cite{falk_wdm}.
Such \emph{warm dense matter} (WDM)~\cite{wdm_book,Dornheim_review} naturally occurs in various astrophysical objects such as giant planet interiors~\cite{Benuzzi_Mounaix_2014,guillot2022giant,drake2018high}, brown dwarfs~\cite{becker} and the outer crust of neutron stars~\cite{neutron_star_envelopes}. Moreover, WDM plays an important role in a variety of technological applications such as inertial confinement fusion (ICF)~\cite{Betti2016}, where the fuel capsule has to traverse the WDM regime in a controlled way to reach ignition ($\rho=500$~g/cc, $T=5\times 10^7$~K)~\cite{hu_ICF}.
The recent breakthrough at the National Ignition Facility (NIF) to reach ignition~\cite{PRL.132.065102} has further substantiated the importance of accurately diagnosing WDM states.
In the laboratory, these extreme states can be created in the laboratory using a variety of experimental techniques~\cite{falk_wdm}, and XRTS is often used to infer a-priori unknown system parameters such as the mass density $\rho$, temperature $T$, and ionization state $Z$; see, e.g., Ref.~\cite{Plagemann2015}. These properties can then be used for physical considerations, to inform equation-of-state tables~\cite{Falk_HEDP_2012,Falk_PRL_2014,Gaffney2018}, and to benchmark integrated multi-scale simulations such as radiation hydrodynamics~\cite{pomraning2005equations}.

In practice, the measured XRTS intensity can be accurately expressed as~\cite{gawne2024effectsmosaiccrystalinstrument}
\begin{eqnarray}\label{eq:convolution}
    I(\mathbf{q},E) = S_{ee}(\mathbf{q},E) \circledast R(E)\ ,
\end{eqnarray}
where $S_{ee}(\mathbf{q},E)$ denotes the electronic dynamic structure factor (DSF) that describes the probed system, and $R(E)$ the combined source-and-instrument function (SIF) that takes into account the shape of the x-ray source and effects of the detector~\cite{gawne2024effectsmosaiccrystalinstrument}.
We note that a deconvolution solving Eq.~(\ref{eq:convolution}) for $S_{ee}(\mathbf{q},E)$ constitutes a major challenge considering the uncertainties of the SIF and the experimental error bars~\cite{Hoell2007,Fortmann2009}.
Therefore, the usual way to interpret the measured XRTS intensity has been to construct a forward model for $S_{ee}(\mathbf{q},E)$, convolve it with $R(E)$, and then fit the resulting trial intensity to the LHS~of Eq.~(\ref{eq:convolution}) where the unknown system properties are treated as free parameters.
Naturally, the thus inferred information depends on the employed forward model for $S_{ee}(\mathbf{q},E)$. Although advances with \textit{ab-initio} methods have been made~\cite{dynamic2}, these models are usually based on a number of assumptions such as the possibility to distinguish between \emph{bound} and \emph{free} electrons as proposed within the popular Chihara approach~\cite{Chihara_1987,Hoell2007,Glenzer_PRL_2007,Lee_PRL_2009,HJLee2010,TDoeppner2010,Fortmann2009,siegfried_review,Gregori_PRE_2003,johnson2012thomson,Souza2014,Mattern2012,kraus_xrts,Tilo_Nature_2023,boehme2023evidence}.

\begin{figure*}\centering
\includegraphics[width=0.99\textwidth]{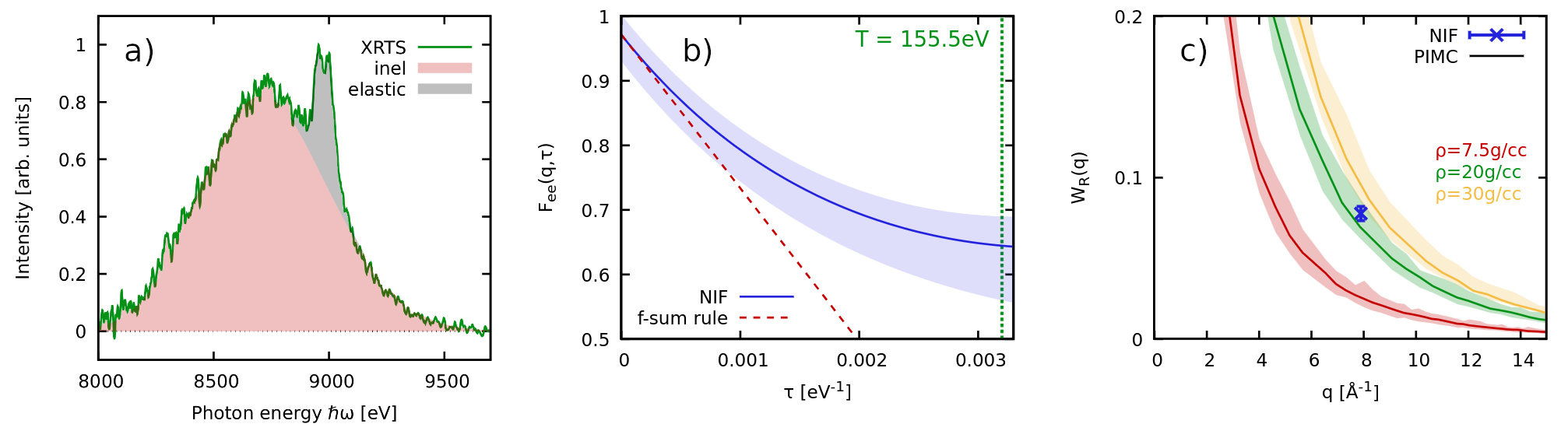}
\caption{\label{fig:WR_Be} Left: XRTS measurement on strongly compressed Be at the NIF~\cite{Tilo_Nature_2023} (green) and its decomposition into elastic (grey) and inelastic (red) contributions. Center: Determination of the normalization $S_{ee}(\mathbf{q})=F_{ee}(\mathbf{q},0)$ from the f-sum rule~\cite{dornheim2023xray}, cf.~Eq.~(\ref{eq:fsum}). Right: Wavenumber dependence of the Rayleigh weight $W_R(\mathbf{q})$, comparing the experimental data point [Eq.~(\ref{eq:final})] with \emph{ab initio} PIMC simulations~\cite{Dornheim_Science_2024}; the solid lines have been obtained for $T=155.5\,$eV and the shaded areas correspond to the uncertainty range of $\pm15\,$eV.
}
\end{figure*}

Very recently, it has been proposed to instead consider the two-sided Laplace transform of the DSF~\cite{Dornheim_T_2022,Dornheim_T_follow_up,Dornheim_review},
\begin{eqnarray}
   F_{ee}(\mathbf{q},\tau) = \mathcal{L}\left[S_{ee}(\mathbf{q},E)\right] = \int_{-\infty}^\infty \textnormal{d}E\ S_{ee}(\mathbf{q},E)\ e^{-\tau E}\ ,
\end{eqnarray}
which is directly related to the imaginary-time density--density correlation function (ITCF) $F_{ee}(\mathbf{q},\tau)$. The latter naturally emerges in Feynman's path-integral picture of statistical mechanics~\cite{kleinert2009path} and, from a physical perspective, contains the same information as $S_{ee}(\mathbf{q},E)$, although in an unfamiliar representation~\cite{Dornheim_MRE_2023,Dornheim_PTR_2023}. It corresponds to the usual intermediate scattering function $F(\mathbf{q},t)=\braket{\hat{n}(\mathbf{q},t)\hat{n}(-\mathbf{q},0)}$ with an imaginary time argument $t=-i\hbar\tau$ with $\tau\in[0,\beta]$ and $\beta=1/k_\textnormal{B}T$ the inverse temperature.
A key advantage of working in the Laplace domain is given by the convolution theorem
\begin{eqnarray}\label{eq:convolution_theorem}
    \mathcal{L}\left[S_{ee}(\mathbf{q},E)\right] = \frac{\mathcal{L}\left[S_{ee}(\mathbf{q},E)\circledast R(E)\right]}{\mathcal{L}\left[R(E)\right]}\ ,
\end{eqnarray}
which is remarkably stable with respect to experimental noise~\cite{Dornheim_T_follow_up,dornheim2023xray}. Eq.~(\ref{eq:convolution_theorem}) thus gives one direct access to $F_{ee}(\mathbf{q},\tau)$ from the experimental observation.
As a first application, Dornheim \emph{et al.}~\cite{Dornheim_T_2022} have suggested to consider the imaginary-time version of the detailed balance relation $F_{ee}(\mathbf{q},\tau)=F_{ee}(\mathbf{q},\beta-\tau)$, which gives one model-free access to the temperature of arbitrarily complex systems~\cite{Dornheim_T_2022,Dornheim_T_follow_up,boehme2023evidence,Schoerner_PRE_2023}. A second practical application of the ITCF in the context of XRTS diagnostics is given by the f-sum rule, which relates the first $\tau$-derivative of $F_{ee}(\mathbf{q},\tau)$ to the momentum transfer $q=|\mathbf{q}|$ that follows from the scattering angle $\theta$~\cite{dornheim2023xray,Dornheim_moments_2023}, see Eq.~(\ref{eq:fsum}), with $m_e$ the electron mass. In this way, one can infer the normalization of the measured intensity, which is given by the electronic static structure factor $S_{ee}(\mathbf{q})$---the Fourier transform of the electron--electron pair correlation function $g_{ee}(\mathbf{r})$.
Finally, we mention the recent idea by Vorberger \emph{et al.}~\cite{Vorberger_PLA_2024}, who have proposed to utilize Eq.~(\ref{eq:convolution_theorem}) to 
quantify the degree of electronic non-equilibrium in the probed system.

In the present work, we extend these efforts towards a model-free diagnostics of XRTS measurements by extracting the Rayleigh weight $W_R(\mathbf{q})=S_{eI}^2(\mathbf{q})/S_{II}(\mathbf{q})$ [where $S_{eI}(\mathbf{q})$ and $S_{II}(\mathbf{q})$ are the electron--ion and ion--ion static structure factor], which describes the degree of electronic localization around the ions~\cite{Vorberger_PRE_2015,Ma_PRL_2013},  directly from the experimental observation.
Our idea is based on a combination of the f-sum rule with the ratio of elastic to inelastic scattering $r(\mathbf{q})$, cf.~Eq.~(\ref{eq:ratio}) below. Therefore, it is generally available and even extends to non-equilibrium situations. As a practical example, we consider an XRTS measurement on strongly compressed Be that has been carried out at the NIF~\cite{Tilo_Nature_2023}. 
The proposed direct inference of $W_R(\mathbf{q})$ helps to further constrain forward models for the interpretation of XRTS measurements, which is very important in its own right. In addition, we expect $W_R(\mathbf{q})$ to be a valuable diagnostic tool.
For example, one might first infer the temperature $T$ from a given XRTS signal using the model-free ITCF method~\cite{Dornheim_T_2022} and then carry out a set of \emph{ab intio} calculations for $W_R(\mathbf{q})$ over a reasonable interval of densities $\rho$; here, we use highly accurate path integral Monte Carlo (PIMC) simulations using the set-up described in Ref.~\cite{Dornheim_Science_2024} and density functional theory molecular dynamics (DFT-MD) simulations by Bethkenhagen \emph{et al.}~\cite{Bethkenhagen}. We find excellent agreement between both methods in $W_R(\mathbf{q})$, whereas computationally less intensive average-atom models~\cite{SHWI07,Souza2014} and parameterized Chihara fits \cite{Gregori_PRE_2003,siegfried_review} deviate. This is of direct practical consequence for the interpretation of the experimental signal and suggests a significantly lower density of $\rho=(22\pm2)$~g/cc (see also Ref.~\cite{Dornheim_Science_2024}) compared to the Chihara based estimate of $\rho=(34\pm4)$~g/cc by D\"oppner \emph{et al.}~\cite{Tilo_Nature_2023}.

From a methodological perspective, 
we note that $W_R(\mathbf{q})$ constitutes a perfect observable for DFT-MD, 
as it only involves the static single-electron density distribution $n_e(\mathbf{r})$ [and the ion density distribution and correlation function, which are both easily accessible within MD]. This is in contrast to time-dependent DFT (TD-DFT) calculations that, in practice, contain additional approximations such as the unknown dynamic exchange--correlation kernel in the case of linear-response TD-DFT~\cite{dynamic2,Moldabekov_JCTC_2023,Moldabekov_PRR_2023,Schoerner_PRE_2023}.
Finally, we mention the possibility to use experimental results for $W_R(\mathbf{q})$ as a rigorous benchmark for simulations and theoretical models in situations where the density and temperature are already known by other means.


\textbf{Idea.} In general, when we can distinguish an elastic peak in an XRTS signal, we can split the DSF into a quasi-elastic contribution $S_\textnormal{el}(\mathbf{q},E)$ 
described by the Rayleigh weight, and an inelastic part $S_\textnormal{inel}(\mathbf{q},E)$ [cf.~Fig.~\ref{fig:WR_Be}a)],
\begin{eqnarray}\label{eq:sum}
    S_{ee}(\mathbf{q},E) &=& \underbrace{W_R(\mathbf{q})\delta(E)}_{S_\textnormal{el}(\mathbf{q},E)}  + S_\textnormal{inel}(\mathbf{q},E) \ .
\end{eqnarray}
Here, the quasi-elastic nature of the first part is due to the substantially longer ionic time-scales and localization of bound and nearly-bound electrons around the ions. In practice, treating $S_\textnormal{el}(\mathbf{q},E)$ as a delta distribution is appropriate if the SIF $R(E)$ is significantly broader than the actual ion feature. This is usually the case except for dedicated experiments that aim to resolve ionic energy scales ~\cite{Descamps2020,White_PRR_2024}.
Within the chemical picture that assumes a decomposition into effectively \emph{bound} and \emph{free} electrons~\cite{siegfried_review,Gregori_PRE_2003,kraus_xrts,Tilo_Nature_2023}, the inelastic part $S_\textnormal{inel}(\mathbf{q},E)$ consists of both transitions between bound and free states (and their reverse process~\cite{boehme2023evidence}) and transitions between two free electronic states. The Rayleigh peak, too, 
is informed by  both free electrons (screening cloud) and bound electrons (form factor), and constitutes an indispensable ingredient to Chihara models~\cite{Chihara_1987,Gregori_PRE_2003,Souza2014} that are widely used for the interpretation of XRTS experiments with WDM~\cite{boehme2023evidence,Tilo_Nature_2023,kraus_xrts,siegfried_review}.
However, it is important to note that we do not have to make such an approximate distinction between bound and free electrons in the present work as the computation of $W_R(q)$ only involves correlation functions between all electrons and ions, i.e., $S_{eI}(q)$ and $S_{II}(q)$.


Let us next consider the ratio of the elastic and inelastic contributions
\begin{eqnarray}\label{eq:ratio}
    r(\mathbf{q}) = \frac{\int_{-\infty}^\infty \textnormal{d}\omega\ S_\textnormal{el}(\mathbf{q},\omega)}{\int_{-\infty}^\infty \textnormal{d}\omega\ S_\textnormal{inel}(\mathbf{q},\omega)} = \frac{W_R(\mathbf{q})}{S_{ee}(\mathbf{q})-W_R(\mathbf{q})}\ ,
\end{eqnarray}
that constitutes a standard observable in XRTS experiments~\cite{Tilo_Nature_2023}, see, e.g., Fig.~\ref{fig:WR_Be}a).
Here $S_{ee}(\mathbf{q})=F_{ee}(\mathbf{q},0)$ denotes the aforementioned electronic static structure factor that we can also directly extract from the XRTS signal using the f-sum rule applied in the imaginary-time domain as has been explained in detail in the recent Ref.~\cite{dornheim2023xray}, see also Eq.~(\ref{eq:fsum}) below. 
Solving Eq.~(\ref{eq:ratio}) for the Rayleigh weight then gives
\begin{eqnarray}\label{eq:final}
    W_R(\mathbf{q}) = \frac{S_{ee}(\mathbf{q})}{1+r^{-1}(\mathbf{q})}\ .
\end{eqnarray}

\begin{figure}\centering
\includegraphics[width=0.392\textwidth]{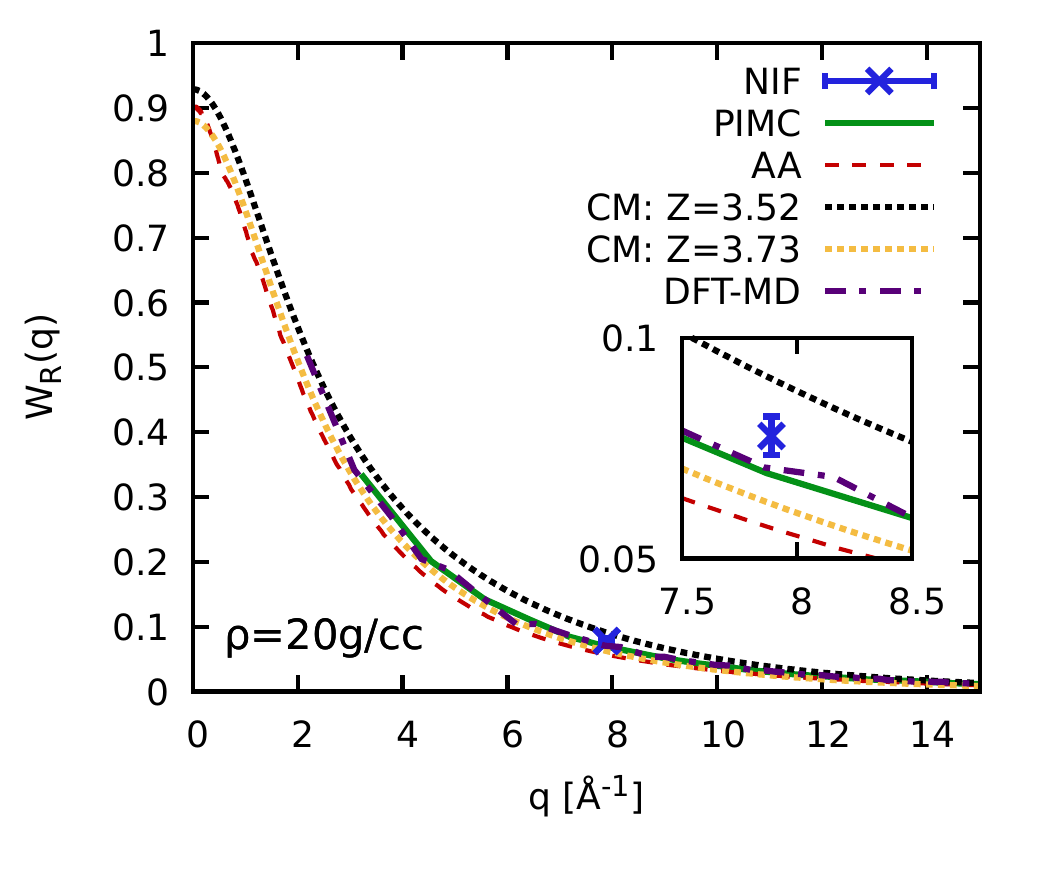}\\\vspace{-1.06cm}
\includegraphics[width=0.392\textwidth]{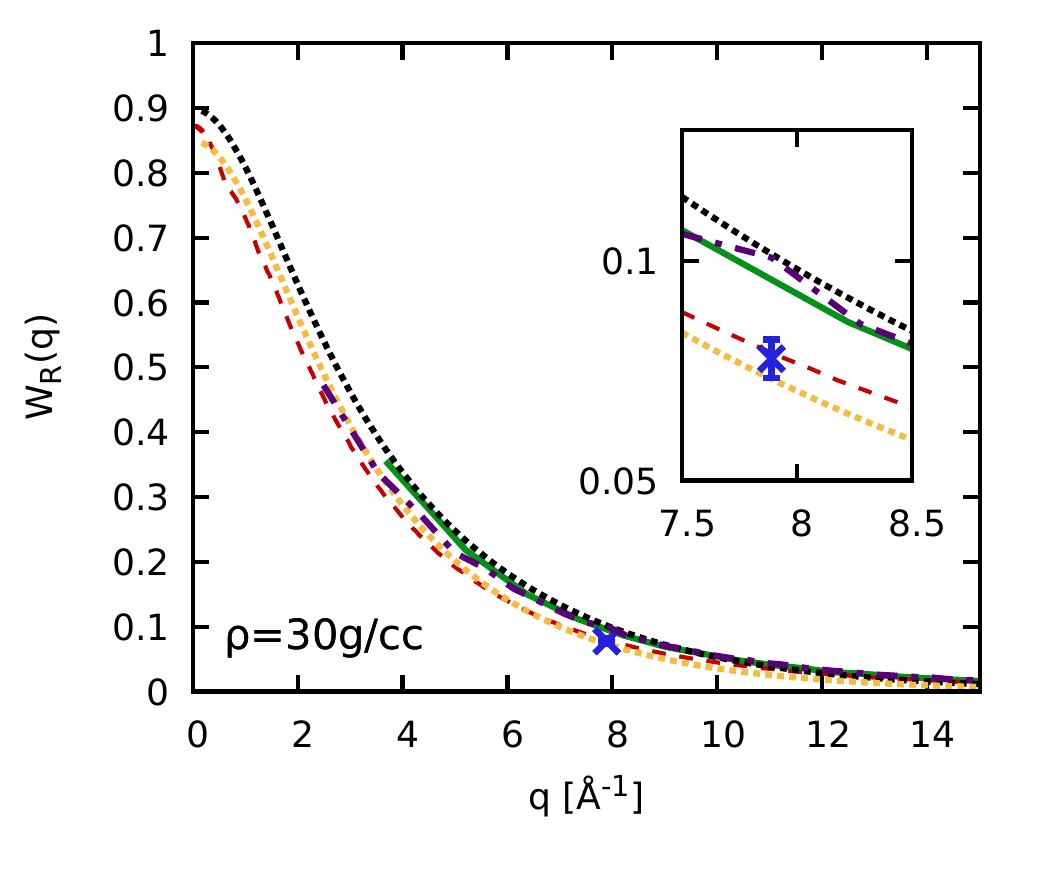}
\caption{\label{fig:model} Comparison of simulation results for the Rayleigh weight at $\rho=20\,$g/cc [top] and $\rho=30\,$g/cc [bottom]. Blue cross: experiment~\cite{Tilo_Nature_2023}; solid green: PIMC; dashed red: average atom model; black and yellow dotted: chemical model with ionization states of $Z=3.52$ and $Z=3.73$; dash-dotted purple: DFT-MD~\cite{Bethkenhagen}. PIMC and CM have been computed for $T=155.5\,$eV, DFT-MD and AA for $T=150\,$eV.
}
\end{figure}

\begin{figure}\centering
\includegraphics[width=0.4\textwidth]{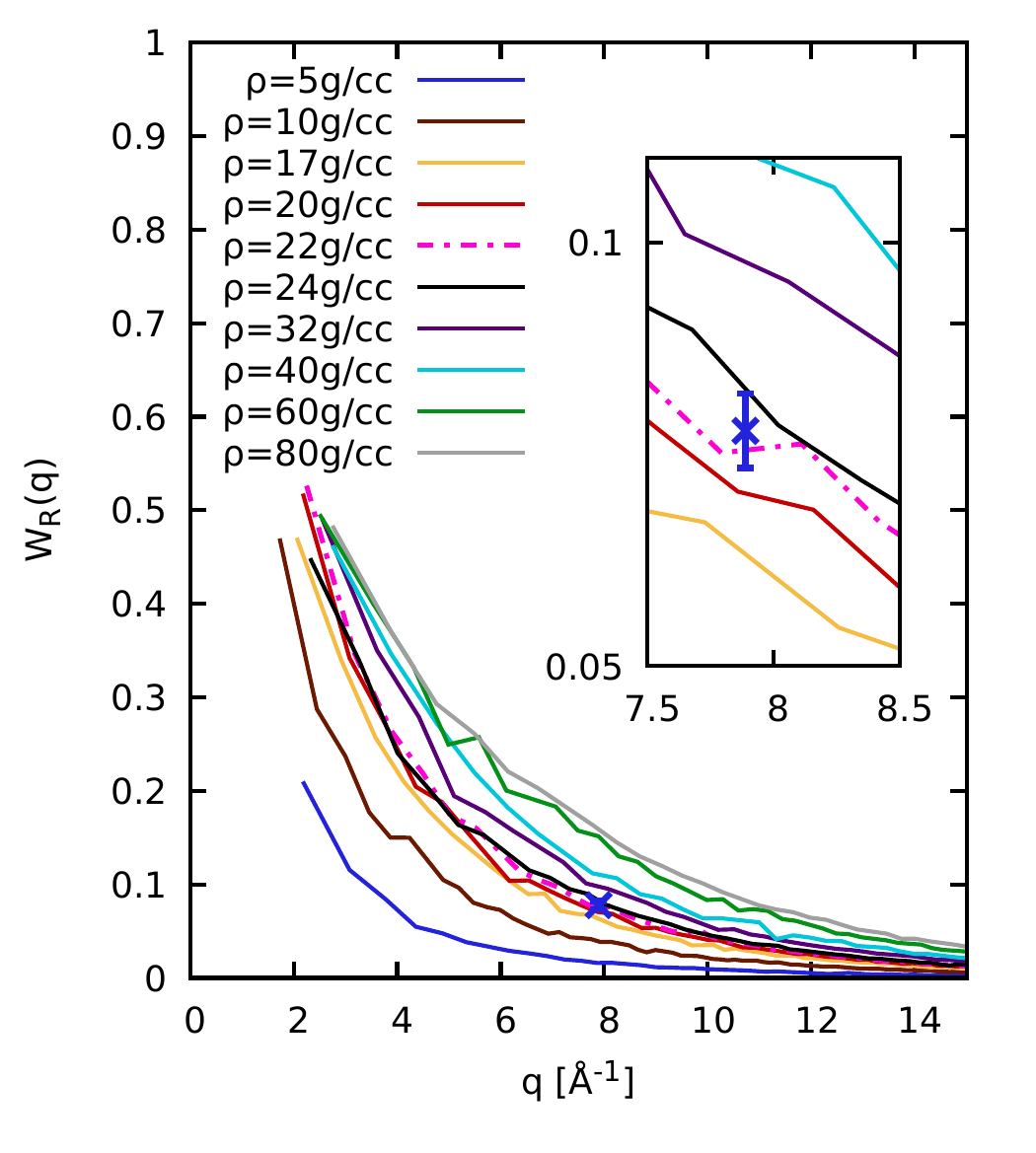}
\caption{\label{fig:DFT} Curves: DFT-MD results for $W_R(\mathbf{q})$ at $T=150\,$eV for different  mass densities $\rho$. The experimental measurement (blue cross) at the NIF~\cite{Tilo_Nature_2023} is associated with a density of $\rho=22\pm2\,$g/cc [dash-dotted pink], see also the inset showing a magnified segment.
}
\end{figure}

\textbf{Results.} In Fig.~\ref{fig:WR_Be}, we apply our new idea to an XRTS data set that has been recently obtained at the NIF by D\"oppner \emph{et al.}~\cite{Tilo_Nature_2023}.
Panel a) shows the full XRTS intensity [solid green, cf.~Eq.~(\ref{eq:convolution})], which can easily be decomposed into its elastic (grey) and inelastic (red) contribution if the SIF $R(E)$ is known; this is indeed the case for the backlighter x-ray source utilized at the NIF~\cite{MacDonald_POP_2022}.
The second ingredient to Eq.~(\ref{eq:final}) is given by the electronic static structure factor $S_{ee}(\mathbf{q})$, which we obtain from the Laplace transform of the XRTS signal via the f-sum rule. This procedure is illustrated in Fig.~\ref{fig:WR_Be}b), where we show the deconvolved $F_{ee}(\mathbf{q},\tau)$ as a function of $\tau$. In the imaginary-time domain, the f-sum rule is given by~\cite{Dornheim_moments_2023,Dornheim_MRE_2023,dornheim2023xray}
\begin{eqnarray}\label{eq:fsum}
  \left.  \frac{\partial}{\partial\tau} F_{ee}(\mathbf{q},\tau)\right|_{\tau=0} = - \frac{\hbar^2 \mathbf{q}^2}{2m_e}\ ;
\end{eqnarray}
matching Eq.~(\ref{eq:fsum}) [dashed red line] with the LHS.~of Eq.~(\ref{eq:convolution_theorem}) [solid blue line] around $\tau=0$ then determines the a-priori unknown normalization constant and, in turn, $S_{ee}(\mathbf{q})=F_{ee}(\mathbf{q},0)$.
Inserting $r(\mathbf{q})$ and $S_{ee}(\mathbf{q})$
into Eq.~(\ref{eq:final}) gives our final, model-free estimate of the Rayleigh weight, which is shown as the blue cross in Fig.~\ref{fig:WR_Be}c) at the corresponding experimental wavenumber of $q=7.89\,$\AA$^{-1}$. 
The red, green and yellow solid curves in the panel show 
\emph{ab initio} PIMC results that have been obtained for $N_\textnormal{Be}=10$ Be atoms using the set-up described in the recent Refs.~\cite{Dornheim_Science_2024,Dornheim_JCP_2024} at the temperature of $T=155.5$~eV that has been inferred from the symmetry of the ITCF; see the appendix of Ref.~\cite{Dornheim_Science_2024} for additional details. 
In addition, the associated colored areas correspond to PIMC simulations for $T=(155.5\pm15)$~eV and indicate the uncertainty in the Rayleigh weight at a given density due to the uncertainty in the inferred temperature.
Fig.~\ref{fig:WR_Be} thus indicates a mass density of $\rho\approx20$~g/cc, whereas the nominal result of $\rho=(34\pm4)$~g/cc reported in the original Ref.~\cite{Tilo_Nature_2023} is ruled out.


For further insight into the sensitivity of the interpretation of a given XRTS measurement to the employed model or simulation technique, we compare a number of independent methods in Fig.~\ref{fig:model}, where the top and bottom panels correspond to $\rho=20$~g/cc and $\rho=30$~g/cc, respectively. For both cases, we find very good agreement between PIMC (solid green) and DFT-MD simulations by Bethkenhagen \emph{et al.}~\cite{Bethkenhagen} (dash-dotted purple). The computationally cheaper average-atom model~\cite{SHWI07,Souza2014} (dashed red) is in qualitative, though not quantitative agreement with the \emph{ab initio} data sets. Instead, it agrees with the experimental data point for $\rho=30$~g/cc. This 
highlights the sensitivity of equation-of-state measurements to the proper treatment of electronic correlations and other many-body effects in the analysis of the experimental observation.
Finally, we include two chemical models for the hypothetical ionization degrees of $Z=3.52$ (dotted black) and $Z=3.73$ (dotted yellow) for both densities, see the Supplemental Material for additional details~\cite{supplement}. Evidently, the inferred density strongly depends on the inferred ionization state, making the invocation of additional constraints desirable. In Ref.~\cite{Tilo_Nature_2023}, D\"oppner \emph{et al.}~have used the Saha equation in combination with a semi-empirical form factor lowering model for this purpose, leading to the nominal parameters of $\rho=(34\pm4)$~g/cc and $Z=3.4\pm0.1$.
A systematic investigation of the origin of this discrepancy between their chemical model and the \emph{ab initio} data sets from PIMC and DFT-MD is beyond the scope of the present work and will be pursued in a dedicated future study.


Let us conclude with a systematic analysis of the dependence of $W_R(\mathbf{q})$ on the density, as shown in Fig.~\ref{fig:DFT} based on extensive DFT-MD results for $\rho=(5-80)$~g/cc at $T=150$~eV.
The main trend is given by the systematic increase of the Rayleigh weight with $\rho$, which is to a large degree due to the different length scales in the system; the effect almost vanishes when one adjusts the $x$-axis by the Fermi wavenumber $q_\textnormal{F}\sim1/\rho^{1/3}$.
The inset shows a magnified segment around the experimental data point, with the latter being located between the DFT-MD results for $\rho=20$~g/cc (red) and $\rho=24$~g/cc (black); this leads to our final estimate for the density of $\rho=(22\pm2)$~g/cc [dash-dotted pink].

\textbf{Summary and Discussion.} We have presented a new approach for the model-free extraction of the Rayleigh weight $W_R(\mathbf{q})$ from XRTS measurements. Most importantly, $W_R(\mathbf{q})$ constitutes an important measure for the electronic localization around the ions, which is related to multiple possible definitions of the average ion charge \cite{murillo2013partial} and interesting in its own right. 
In addition, $W_R(\mathbf{q})$ is of direct practical value for XRTS diagnostics, as we have demonstrated in detail on a recent experiment with strongly compressed beryllium at the NIF~\cite{Tilo_Nature_2023}. In particular, we propose to first infer the temperature $T$ from the model-free ITCF thermometry approach~\cite{Dornheim_T_2022,Dornheim_T_follow_up}, and to subsequently match the experimental result for $W_R(\mathbf{q})$ with simulation results over a reasonable range of densities.
We note that $W_R(\mathbf{q})$ is a particularly suitable observable for DFT-MD simulations, as it does not involve any dynamic information such as the \textit{a-priori} unknown dynamic exchange--correlation kernel. We find excellent agreement between DFT-MD and 
PIMC reference data; this is encouraging since direct PIMC simulations are still limited to low-$Z$ materials at moderate to high temperatures~\cite{dornheim_sign_problem,bonitz2024principles}, whereas DFT-MD and TD-DFT \cite{dynamic2,Gawne_PRB_2024} are more broadly applicable. 
An additional advantage of $W_R(\mathbf{q})$ over alternative observables such as the ratio of elastic to inelastic scattering $r(\mathbf{q})$ is that the former does not require any explicit information about the electronic static structure factor $S_{ee}(\mathbf{q})$, which is notoriously difficult for DFT-based methodologies.

In practice, we infer a mass density of $\rho=(22\pm2)$~g/cc from the beryllium data set using either PIMC or DFT-MD simulations, which is significantly lower than the nominal value of $\rho=(34\pm4)$~g/cc that has been reported in the original publication based on a chemical Chihara model; note that the latter must take into account a proper multi-component treatment of differently charged ions~\cite{wuensch_epl_94}. 
Moreover, we find that the computationally more efficient average-atom model tends to overestimate the density, whereas chemical models generally require additional constraints for parameters such as the ionization degree.
Our study thus clearly highlights the importance of an accurate treatment of quantum many-body effects for XRTS based equation-of-state measurements even at relatively high temperatures.


This work opens up new possibilities for the study of warm dense matter and beyond. An important point for future research is given by the sensitivity of the inferred Rayleigh weight to the SIF $R(E)$, which is usually modelled for backlighter set-ups~\cite{MacDonald_POP_2022}, but can be known with high precision at modern XFEL facilities~\cite{gawne2024effectsmosaiccrystalinstrument}. Indeed, recent advances in high-resolution XRTS measurements~\cite{Gawne_PRB_2024} at the European XFEL will likely facilitate the application of the model-free ITCF thermometry technique even at moderate temperatures of $T\sim1\,$eV, which are of relevance for both planetary and material science; the current model-free framework for the extraction of $W_R(\mathbf{q})$ is applicable at any temperature. In combination, these two methods will allow for accurate equation-of-state measurements in previously inaccessible regimes.
Finally, we mention the intriguing possibility of performing an XRTS experiment on an isochorically heated sample with an appropriate delay between pump and probe to ensure proper equilibration. Since both $T$ and $\rho$ would be well known in such a scenario, a corresponding measurement of $W_R(\mathbf{q})$ using the present framework would constitute a truly unambiguous reference data set for the rigorous benchmarking of \emph{ab initio} simulations and chemical models alike.




\section*{Acknowledgments}
This work was partially supported by the Center for Advanced Systems Understanding (CASUS), financed by Germany’s Federal Ministry of Education and Research (BMBF) and the Saxon state government out of the State budget approved by the Saxon State Parliament.
This work has received funding from the European Union's Just Transition Fund (JTF) within the project \emph{R\"ontgenlaser-Optimierung der Laserfusion} (ROLF), contract number 5086999001, co-financed by the Saxon state government out of the State budget approved by the Saxon State Parliament.
This work has received funding from the European Research Council (ERC) under the European Union’s Horizon 2022 research and innovation programme
(Grant agreement No. 101076233, "PREXTREME"). 
This work was partially supported by the German Research Foundation (DFG) within the Research Unit FOR~2440.
Views and opinions expressed are however those of the authors only and do not necessarily reflect those of the European Union or the European Research Council Executive Agency. Neither the European Union nor the granting authority can be held responsible for them. Computations were performed on a Bull Cluster at the Center for Information Services and High-Performance Computing (ZIH) at Technische Universit\"at Dresden and at the Norddeutscher Verbund f\"ur Hoch- und H\"ochstleistungsrechnen (HLRN) under grants mvp00018 and mvp00024.
SH was supported by Sandia National Laboratories, a multi-mission laboratory managed and operated by National Technology and Engineering Solutions of Sandia, LLC, a wholly owned subsidiary of Honeywell International, Inc., for DOE's National Nuclear Security Administration under contract DE-NA0003525.
This paper describes objective technical results and analysis.
Any subjective views or opinions that might be expressed in the paper do not necessarily represent the views of the U.S. Department of Energy or the United States Government.

\appendix

\newpage

\LARGE{Supplemental Material}
\normalsize

\section{Uncertainty estimate of NIF data point}\label{sec:NIF}

As specified in Eq.~(6) of the main text, the model-free extraction of the Rayleigh weight from the measured XRTS intensity is based on two ingredients, each of which contributes to the total uncertainty of the final result for $W_R(\mathbf{q})$: i) the electron--electron static structure factor $S_{ee}(\mathbf{q})$ and ii) the ratio of elastic to inelastic scattering $r(\mathbf{q})$.

i) In principle, evaluating the two-sided Laplace transform would require an infinite integration range, whereas the spectral window in the experiment is always finite. In practice, we thus define the symmetrically truncated Laplace transform
\begin{eqnarray}\label{eq:truncated}
    \mathcal{L}\left[S_{ee}(\mathbf{q},E)\right] = \int_{-x}^x\textnormal{d}E\ S_{ee}(\mathbf{q},E)\ e^{-\tau E}\ ,
\end{eqnarray}
and investigate the convergence with respect to $x$, see Ref.~\cite{dornheim2023xray} for a detailed description of its implications for the determination of $S_{ee}(\mathbf{q})$ from the f-sum rule, i.e., from the first derivative of $F_{ee}(\mathbf{q},\tau)$ with respect to $\tau$ around $\tau=0$.
For the XRTS data set of strongly compressed beryllium that is considered in the present work, a corresponding convergence analysis is shown in Ref.~\cite{Dornheim_Science_2024}, leading to $S_{ee}(\mathbf{q}) = 0.971\pm 0.036$.

ii) To decompose the experimental intensity signal $I(\mathbf{q},E)$ into an elastic and inelastic contribution as it has been depicted in Fig.~1a) of the main text, we only need to assume that $S_\textnormal{inel}(\mathbf{q},E)$ is smooth around the beam energy $E_0=9000\,$eV. In practice, it is convenient to use a Chihara fit onto the observed data for this purpose, which leads to $r(\mathbf{q})=0.087\pm0.004$~\cite{Tilo_Nature_2023}; the corresponding error is obtained from a Gaussian error propagation onto the elastic-to-inelastic ratio from the given fit uncertainties. 
It is important to note that $r(\mathbf{q})$ is to a very high degree independent from a particular forward model for $S_{ee}(\mathbf{q},E)$ as long as the convolution with the SIF fits the experimental signal $I(\mathbf{q},E)$. This is in stark contrast to parameters such as the density, temperature, or charge state.  We also note that while the core 1s electrons are tightly bound and well localized in the present case, other materials at other conditions may have loosely bound valence electrons that increase uncertainties in isolating the elastic contribution.

The final uncertainty of our estimate for $W_R(\mathbf{q})$ then follows from a straightforward error propagation analysis of Eq.~(6), and we find $W_R(\mathbf{q})=0.0778\pm0.0044$.

\section{Range of applicability and limitations of the method}\label{sec:limits}

Being exclusively based on the ratio of elastic-to-inelastic scattering and the static structure factor, the method is applicable to a wide range of conditions. In particular, one only needs to resolve the ITCF $F_{ee}(\mathbf{q},\tau)$ around the vicinity of $\tau=0$. This is in stark contrast to the model-free ITCF thermometry approach discussed in Refs.~\cite{Dornheim_T_2022,Dornheim_T_follow_up}, for which one needs to resolve the ITCF for $\tau \gtrsim \beta/2$. As a consequence, the temperature extraction only works for sufficiently hot systems (i.e., small enough $\beta$), whereas the present approach works well for all temperatures, including ambient conditions~\cite{dornheim2023xray}.

A more serious point of consideration is given by the convolution theorem [Eq.~(3) of the main text]. Most fundamentally, this relation assumes that the SIF $R(E)$ does not depend on the photon energy. A dedicated recent analysis by Gawne \emph{et al.}~\cite{gawne2024effectsmosaiccrystalinstrument}
has shown that this is indeed the case to a high degree for usual experimental energy ranges, and, therefore, can be safely neglected here. Nevertheless, this subtlety might become important if one were to measure the XRTS spectrum over several thousand eV, as it would be required for heavier elements, see below.
A second point concerning Eq.~(3) is given by knowledge of $R(E)$ itself, which is indispensable for both forward-modeling based methodologies~\cite{siegfried_review} as well as the present model-free approach. While $R(E)$ can be inferred at modern XFEL facilities such as the European XFEL with high confidence~\cite{gawne2024effectsmosaiccrystalinstrument}, its characterization for backlighter sources like the one utilized for the XRTS measurement on strongly compressed beryllium at the NIF~\cite{Tilo_Nature_2023} is more difficult~\cite{MacDonald_POP_2022}. Clearly, any uncertainty related to $R(E)$ directly translates to the model-free estimate of $W_R(\mathbf{q})$ using the present approach, and might thus influence the extracted system parameters such as the mass density $\rho$. Hence, we strongly encourage dedicated measurements for the characterization of $R(E)$ for any given experimental set-up if possible.

In addition to these technical considerations, other potential sources of uncertainty concern the probed state of matter itself. A well-known issue related to the experimental realization of warm dense matter states~\cite{falk_wdm} is given by possible non-equilibrium effects. These can potentially be detected without the need for models from the ITCF itself, see the recent Ref.~\cite{Vorberger_PLA_2024}.
A second issue that might be important -- in particular for spherical implosion experiments -- is given by the potential inhomogeity of the compressed sample~\cite{Chapman_POP_2014,Tilo_Nature_2023}, which deserves dedicated attention in future studies. A third concern is related to the treatment of loosely bound or ``hopping'' electronic states, which have some properties of bound states (localization near the ions) and some properties of free states (extensive tails). This concern is related to the Chihara distinction between free and bound electrons \cite{murillo2013partial}.
Finally, we mention the importance of resolving all important spectral features in the XRTS measurement for a given material; this is important both for the determination of $r(\mathbf{q})$ and of $S_{ee}(\mathbf{q})$ (though not for the ITCF thermometry approach~\cite{Dornheim_T_follow_up}), and requires increasingly large spectral ranges with increasing nuclear charge.

\section{Path integral Monte Carlo}\label{sec:PIMC}

As written in the main text, the Rayleigh weight is easily computed from the electron--ion and ion--ion static structure factors via~\cite{Vorberger_PRE_2015}
\begin{eqnarray}\label{eq:W_R}
    W_R(\mathbf{q}) = \frac{S_{eI}^2(\mathbf{q})}{S_{II}(\mathbf{q})}\ .
\end{eqnarray}
Both static structure factors are directly accessible to PIMC simulations; we use the approach presented in the recent Refs.~\cite{Dornheim_Science_2024,Dornheim_JCP_2024} using $P=200$ imaginary-time propagators where the divergent electron--ion Coulomb attraction is treated via the pair approximation as described in Refs.~\cite{Bohme_PRE_2023,MILITZER201688}.
All PIMC simulations are carried out with the \texttt{ISHTAR} code~\cite{ISHTAR}, using the canonical adaption of the worm algorithm by Boninsegni \emph{et al.}~\cite{boninsegni1,boninsegni2} that has been introduced in 
Ref.~\cite{Dornheim_PRB_nk_2021}.

\section{Density functional theory molecular dynamics}\label{sec:DFT}

DFT-MD simulations give snapshots of the ion positions in the simulation box as well as the electronic density on a grid in the same supercell. These can be used to calculate the ion-ion structure factor
\begin{eqnarray}\label{eq:Sii_DFT}
    S_{II}(\mathbf{q})=\frac{1}{N_I}\langle \rho_I(\mathbf{q})\rho_I^*(\mathbf{q})\rangle\ ,
\end{eqnarray}
as well as the electron-ion static structure factor
\begin{eqnarray}\label{eq:Sei_DFT}
    S_{eI}(\mathbf{q})=\frac{1}{\sqrt{N_e N_I}}\langle \rho_e(\mathbf{q})\rho_I^*(\mathbf{q})\rangle\ ,
\end{eqnarray}
from which the Rayleigh weight derives~\cite{Vorberger_PRE_2015}.

The DFT-MD simulations along the 150~eV isotherm were performed with the Vienna Ab initio Simulation Package (VASP)~\cite{Kresse1993,Kresse1994,Kresse1996} and were partially taken from Refs.~\cite{Tilo_Nature_2023, Bethkenhagen}. Here, we considered 11 densities ranging from 5~g/cc to 80~g/cc, whereas the calculations were carried out with $8$--$64$ beryllium atoms in the simulation box depending on the simulated density. All electrons were treated explicitly using the Coulomb potential with a cutoff energy of 10~keV. The DFT-MD simulations were typically run for at least 20~000 time steps after equilibration with a time step size between 5~as and 50~as. The ion temperature was controlled by a  Nos\'{e}-Hoover thermostat~\cite{Nose1984}, the Brillouin zone was sampled at the Baldereschi mean value point~\cite{Baldereschi1973}, and we employed the exchange-correlation (XC) functional of Perdew, Burke, and Ernzerhof (PBE)~\cite{Perdew1996}. Note, that the ion density can be calculated in every timestep, while the electron density was only obtained for 15-20 snapshots per DFT-MD simulation. Therefore, the ion-ion structure factor converges much faster than the electron-ion static structure factor.

\begin{figure*}\centering
\includegraphics[width=0.392\textwidth]{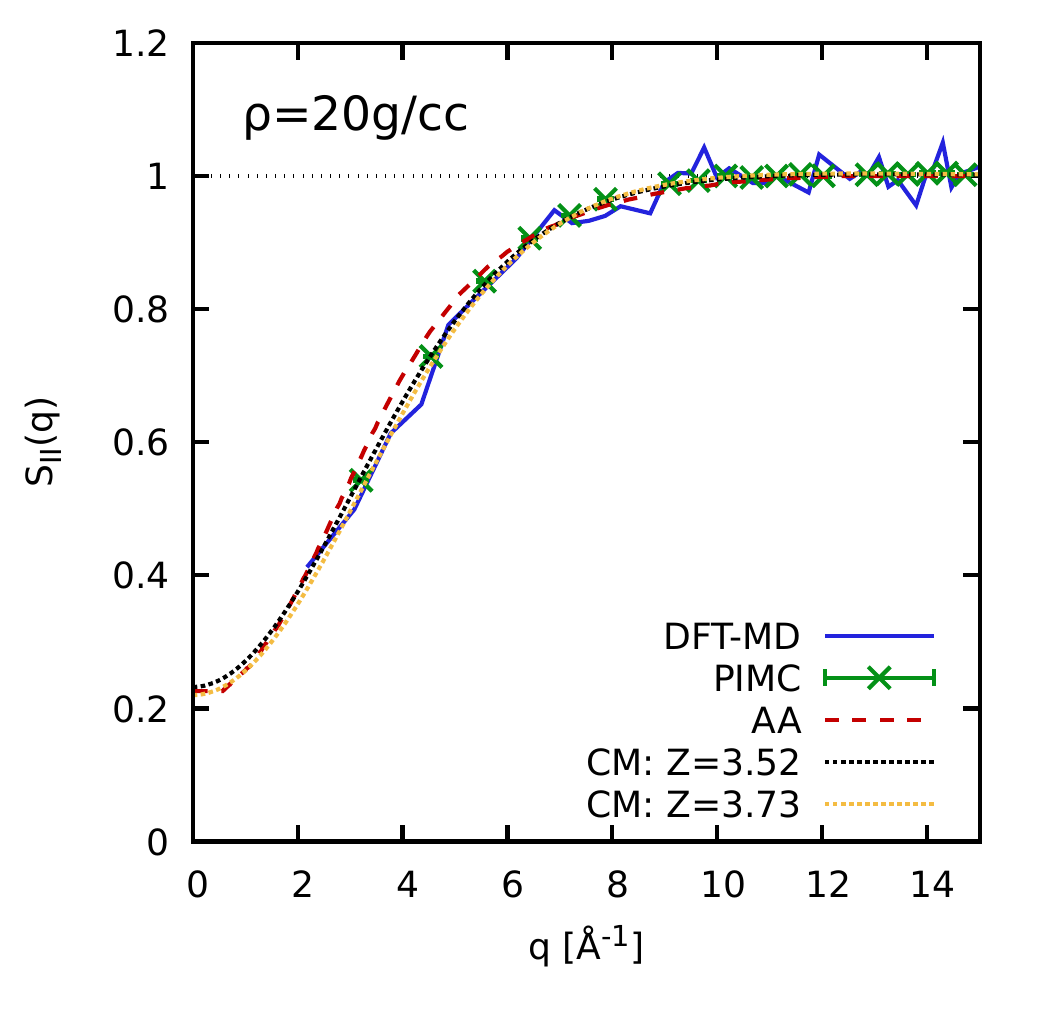}
\includegraphics[width=0.392\textwidth]{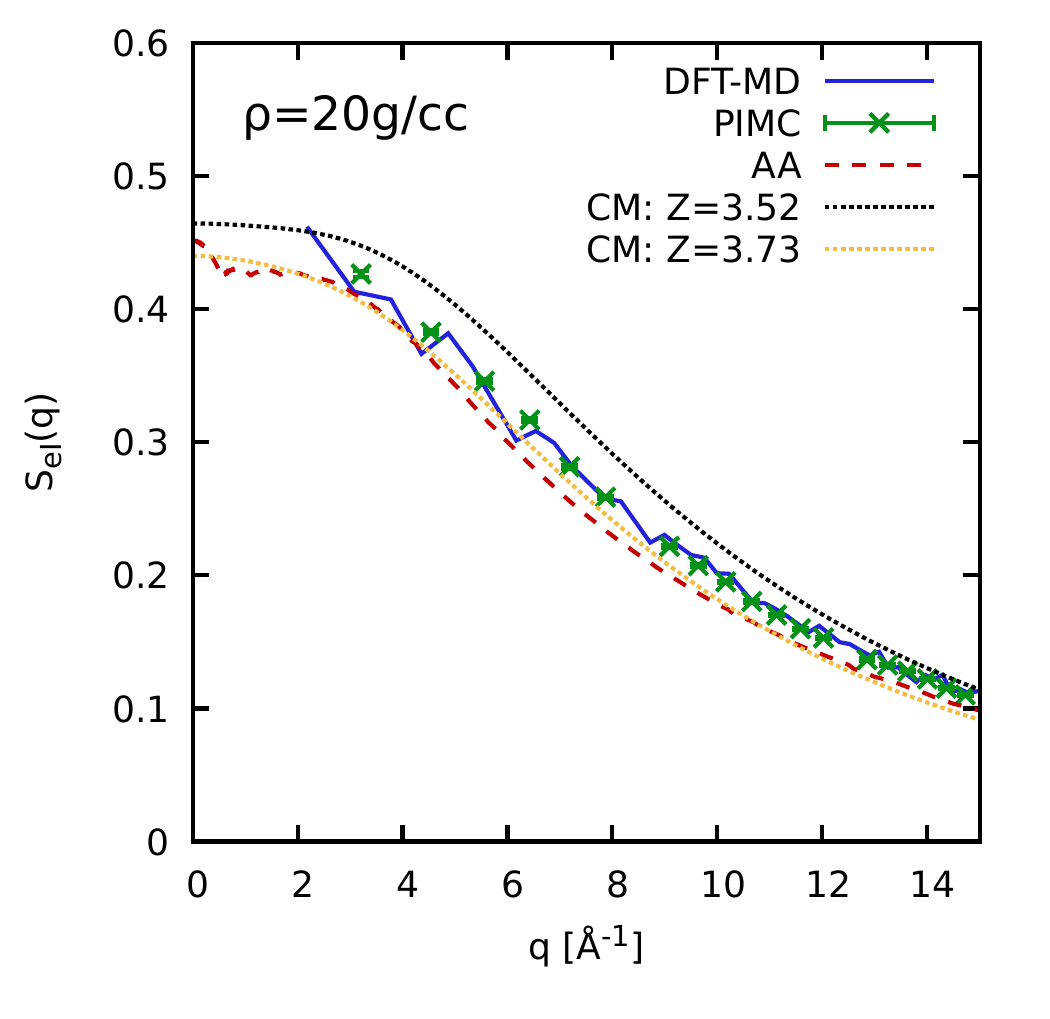}
\caption{\label{fig:SSF} Ion--ion static structure factor $S_{II}(q)$ [left] and electron--ion static structure factor $S_{eI}(q)$ [right] for $\rho=20\,$g/cc and $T\sim150\,$eV. The green crosses and dashed red lines correspond to PIMC (taken from Ref.~\cite{Dornheim_Science_2024}) and average atom, respectively and the yellow (black) dotted curve to the chemical model with $Z=3.73$ ($Z=3.52$).
}
\end{figure*} 

\section{Average atom model}\label{sec:AA}
Average-atom models \cite{Liberman1979,Wilson2006,SHWI07} solve for the real-space electron density around a single ion, using density functional theory to obtain all-electron, spherically averaged densities $n_e^{\mathrm{full}}(r)$ and self-consistent radial potentials within an ion sphere defined by the Wigner-Seitz radius. The model used here is a semi-relativistic version of the Purgatorio code \cite{Wilson2006,SHWI07} that has been extended following Starrett and Saumon \cite{STARRETT201435} to compute the response of the background plasma by solving for the external electron density $n_e^{\mathrm{ext}}(r)$ in the ion sphere without a central nuclear charge. The total form factor $f(q)$ is given by the Fourier transform of the pseudo-atom density $n_e^{\mathrm{PA}}(r)=n_e^{\mathrm{full}}(r)-n_e^{\mathrm{ext}}(r)$. Note that since $\int\,n_e^{\mathrm{PA}}(r)d^3r=Z_n$, with $Z_n$ the nuclear charge, $f(0)=Z_n$. A screening density is defined by subtracting the bound electron density from $n_e^{\mathrm{PA}}(r)$. The transform of the screening density is combined with a response function to generate an ion-ion potential that determines the ion structure factor $S_{II}(q)$. Finally, the Rayleigh weight is obtained via: 

\begin{equation}
   W_R(q)=\frac{f^2(q) S_{II}(q)}{Z_n} \ .
\end{equation}

Here, we use Kohn-Sham LDA for the exchange potential and calculate the ion charge by integrating the free-electron density of states multiplied by the Fermi occupation factor with a self-consistent chemical potential. For Be at 20 g/cc and 150 eV, we find an average ionization of 3.1 and relatively minor ($\approx$10\%) variations of $W_R(q)$ with alternative exchange potentials and definitions of the average ion charge. With this AA model, XRTS spectra can also be directly computed following \cite{johnson2012thomson,Souza2014}. Finally, we note that using the Fourier transform of the average-atom electron density $n_e^{\mathrm{full}}(r)$ for $f(q)$ can give oscillations inconsistent with the $W_R(q)$ of PIMC and DFT-MD.

\section{Chihara models}\label{sec:chem}

The Chihara decomposition \cite{Chihara_1987} is used to separate out contributions from bound and free electrons.
Thus, in this picture, the Rayleigh weight is calculated by considering the ionic contributions.

\begin{figure}\centering
\includegraphics[width=0.392\textwidth]{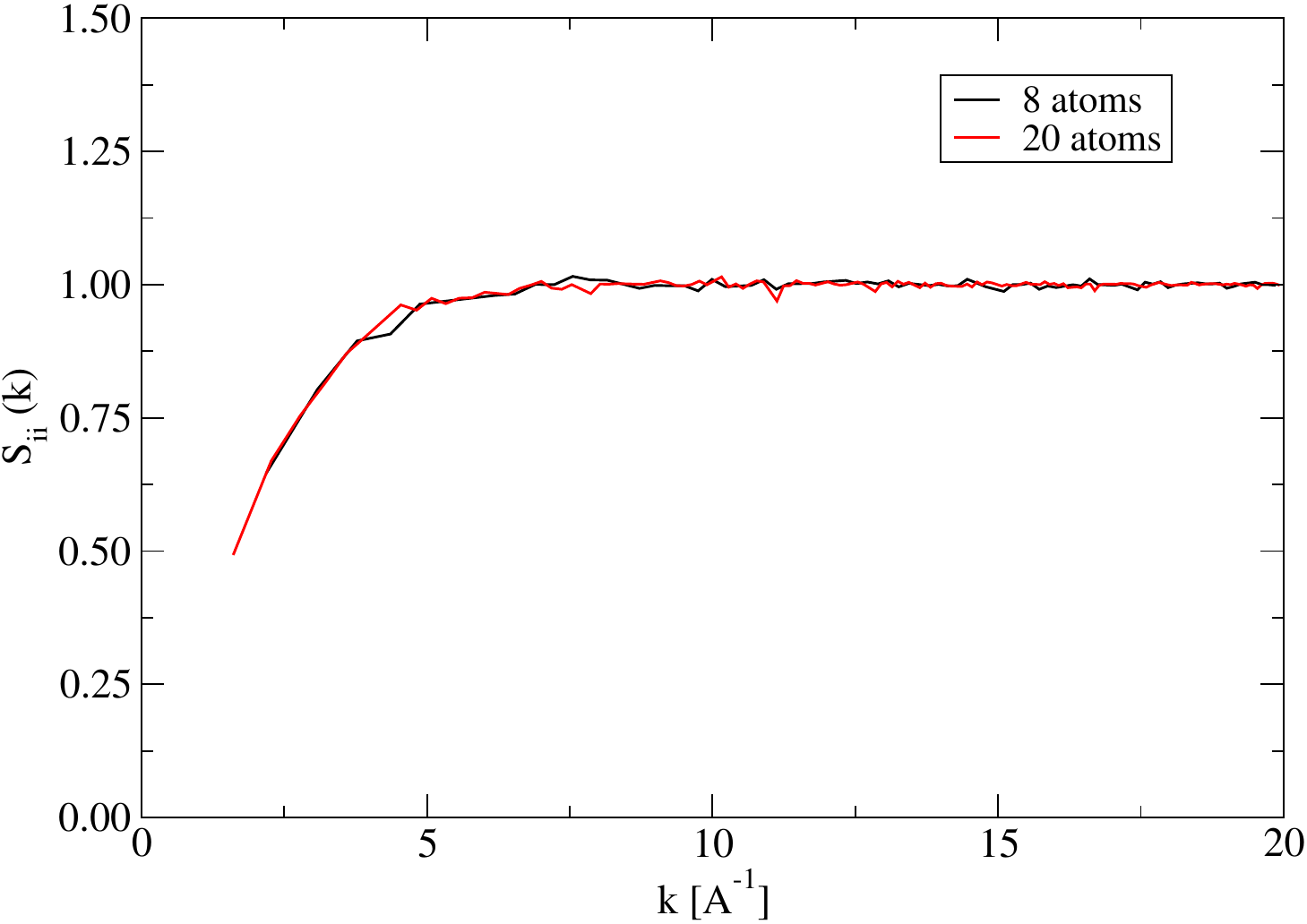}
\caption{\label{fig:SSF_fsc} DFT-MD results for the ion--ion static structure factor $S_{II}(q)$ at $T=150\,$eV and $\rho=5\,$g/cc for $N_{Be}=8$ (black) and $N_{Be}=20$ (red) Be atoms.
}
\end{figure}

More specifically, for a multi-component system with species $a$ and $b$, the Rayleigh weight is calculated as \cite{wuensch_epl_94}: 
\begin{multline}
    W_R (q) = \sum_{a,b} \Bigl\{ \sqrt{x_a x_b} \left[f_{\text{a}}^{\text{ff}} (q) + f_{\text{a}}^{\text{sc}} (q) \right] \\\times \left[f_{\text{b}}^{\text{ff}} (q) + f_{\text{b}}^{\text{sc}} (q) \right] S_{ab} (q) \Bigr\} \,.
\end{multline}
with $q=|\mathbf{q}|$ and $x_a$ being the partial number density fraction of ions of species $a$ given by
\begin{equation}
     x_a = n_a /n_i 
\end{equation}
for the mean number density $n_a$ of species $a$. 
The total number density of ions (with the mean ion mass) in the plasma is 
\begin{equation}
    n_i = \sum_a n_a \,.
\end{equation}

Here, the ionic form factor for a given species, e.g., $a$,  is defined as $f_{\text{a}}^{\text{ff}} (q)$ and $f_{\text{a}}^{\text{sc}} (q)$ denotes the respective screening cloud.
Partial static structure factors $S_{ab}$ account for correlations between the ion species.

A finite wavelength approximation was used for the screening cloud, with an effective Coulomb potential describing the electron-ion interactions.
Form factors were calculated using the screened hydrogenic approximation \cite{pauling_1932}.
The partial static structure factors were calculated using a Debye-H{\"u}ckel effective potential \cite{gericke_pre_2010} within the hypernetted-chain approximation \cite{wuensch_pre_2008}.


\newpage
\section{Static structure factors}

In Fig.~\ref{fig:SSF}, we compose the full Rayleigh weight at $T=150\,$eV and $\rho=20\,$g/cc into its individual components, i.e., the ionic static structure factor $S_{II}(q)$ [left] and the electron--ion static structure factor $S_{eI}(q)$ [right].
Interestingly, we find overall very good agreement between all depicted models in the former, while there are more substantial disagreements in the latter. 
We thus conclude that the accurate description of the electronic localization around the ions poses the main challenge for the accurate description of $W_\textnormal{R}(q)$, whereas ionic correlations are less sensitive to the included microphysics.

Finally, we show DFT-MD results for $S_{II}(q)$ at $T=150\,$eV and $\rho=5\,$g/cc for $N_{Be}=8$ (black) and $N_{Be}=20$ (red) beryllium atoms in Fig.~\ref{fig:SSF_fsc}.
Evidently, there are no finite-size effects except for the different $q$-grids, which is a consequence of momentum quantization in the finize simulation cell.


\bibliography{bibliography}
\end{document}